\documentclass[aps,twocolumn,prl,floats]{revtex4}
\usepackage{amssymb}
\usepackage{graphicx}
\usepackage{dcolumn}
\usepackage{bm}
\usepackage{amsmath}
\usepackage{epsfig}

\begin{document}

\title{The Role of the Gouy Phase in the Coherent Phase Control of the Photoionization
and Photodissociation of Vinyl Chloride}
\author{Vishal J. Barge$^1$}
\author{Zhan Hu$^{1,2}$}
\author{Joyce Willig$^1$}
\author{Robert J. Gordon$^{1*}$}
\affiliation{$^1$Department of Chemistry, University of Illinois at
Chicago, Chicago, IL 60680-7061, USA}
\affiliation{$^2$Institute of Atomic and Molecular Physics, Jilin University, Changchun, P.R. China 130021}

\begin{abstract}
We demonstrate theoretically and experimentally that the Gouy phase
of a focused laser beam may be used to control the photo-induced
reactions of a polyatomic molecule. Quantum mechanical interference
between one- and three-photon excitation of vinyl chloride produces
a small phase lag between the dissociation and ionization channels
on the axis of the molecular beam. Away from the axis, the Gouy
phase introduces a much larger phase lag that agrees quantitatively
with theory without any adjustable parameters.
\end{abstract}

\maketitle

    It is an axiom of quantum mechanics that the probability of an
event may be calculated by adding the probability amplitudes of all
independent paths connecting the initial and final states and then
taking the modulus squared of that sum \cite{Feynman}. Because the
phases of different paths vary with the parameters of the system,
the transition probability displays an oscillatory pattern with
respect to those parameters. Brumer and Shapiro \cite{BS} predicted
that, by manipulating the appropriate parameters, an experimenter
could control the outcome of the event.  Their theory has been
validated experimentally for numerous systems [3-8].

    The most commonly studied control scenario is the
multiphoton excitation of a target by different numbers of photons
in each path. Brumer and Shapiro showed that for the absorption of
$n$ photons of frequency $\omega_m$ and $m$ photons of frequency
$\omega_n$, such that $n\omega_m=m\omega_n$, the probability of
obtaining product $S$ is given by
\begin{equation}
P^S=P^S_m+P^S_n+2P^S_{mn}\cos(\phi_{sp}+\delta^S_{mn}),
 \label{probability}
\end{equation}
where $P^S_m$ is the $n-$photon transition probability, $P^S_n$ is
the $m-$photon probability, and $P^S_{mn}$ is the amplitude of the
interference term \cite{SHB}. The interference term is given
explicitly  by the integral
\begin{equation}
P^S_{mn}e^{i(\phi_{sp}+\delta^S_{mn})}=\int d\hat{k} \langle
g|D^{(m)}|E,S,\hat{k}\rangle \langle E,S,\hat{k}|D^{(n)}|g\rangle,
 \label{Pmn}
\end{equation}
where $|g\rangle$ is the ground state, $|E,S,\hat{k}\rangle$ is the
excited continuum state, $E$ and $\hat{k}$ are the energy and
momentum of the excited state, and  $D^{(j)}$ is the j-photon
transition dipole operator.  The phase of this term consists of a
spatial component, $\phi_{sp},$ which is a property of the radiation
field (contained in  $D^{(j)}$), and a molecular component,
$\delta^S_{mn},$ which depends on the electronic structure of the
target \cite{Seideman}, \cite{JPC}. The molecular phase (also known
as the channel phase) may arise, for example, from coupling of
electronic continua, from a resonant state embedded in the continuum
(both contained in $|E,S,\hat{k}\rangle$) \cite{PRL1}, or from an
intermediate resonant state (contained in $D^{(j)}$) \cite{PRL2}.
Because $\delta^S_{mn}$ is channel-dependent, it is possible to
control the product distribution by manipulating $\phi_{sp}$.

The spatial phase itself has three components,
\begin{eqnarray}
\phi_{sp}&=&(m\phi_n-n\phi_m) +(mk_nz-nk_mz)\nonumber \\
&+&(m-n)\eta(z) , \label{spatial}
\end{eqnarray}
where $\phi_i$ is a constant phase of the electric field, $z$ is the
axial coordinate of the  field, $k_i$ is the wave number,
$\eta(z)=\tan^{-1}(z/z_R)$ is the Gouy phase, and $z_R$ is the
Rayleigh range \cite{Yariv}. The first term in $\phi_{sp}$ is
proportional to the difference between the refractive indices at
frequencies $\omega_m$ and $\omega_n$ \cite{refractive}. The second
term is usually assumed to vanish because of momentum conservation
(although see ref. \cite{Pegarkov} for a possible counter-example).
The Gouy phase shift in the third term results from the increased
phase velocity of a Gaussian beam, as compared with a plane wave, as
it propagates through a focal region \cite{Gouy}, \cite{Siegman},
\cite{Boyd}. More generally, it has been shown that the Gouy phase
results from a spread in the transverse momentum of the focused beam
\cite{Feng}. This phase does not appear in Brumer and Shapiro's
formulation, presumably because it is not explicitly
channel-dependent. Chen and Elliott \cite{ChenElliott} demonstrated
that the modulation of the signal produced by one- and three-photon
ionization of mercury atoms undergoes a $\pi$ phase shift as the
probed region passes through the focal point of the laser beams. In
all previous phase control experiments, the refractive term
($m\phi_n-n\phi_m$) was adjusted experimentally to cancel the
molecular phase for a selected channel, thereby enhancing the yield
from that channel. Here we demonstrate that the Gouy phase may be
exploited to control a branching ratio, even in the absence of a
molecular phase.

\begin{figure}[htbp]
\begin{center}
\vspace*{-0mm} \hspace{2mm} {\hbox{\epsfxsize=80mm
\epsffile{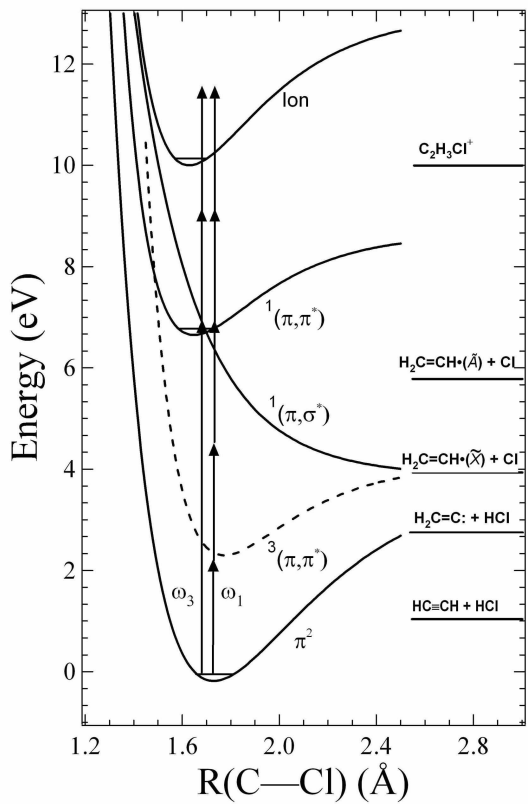}}} \vspace{0mm}
\end{center}
\caption{Schematic slice of the potential energy surfaces of vinyl
chloride, showing the interfering excitation paths.}
\label{Potential Energy}
\end{figure}

     The system we chose to study is the photodissociation and
photoionization of vinyl chloride ($CH_2=CHCl$, $VCl$).  A potential
energy diagram for these reactions is shown in Fig. \ref {Potential
Energy} \cite{PES1}. In our study, three 532 nm photons (at
frequency $\omega_1$) and one 177 nm photon (at frequency
$\omega_3$) are used to excite the molecule to a quasi-bound
$^1\pi,\pi^*$ state.  At this energy level, the molecule can either
predissociate to yield $Cl+C_2H_3$ fragments (among others
\cite{VCl}), or it may absorb two additional 532 nm photons and
ionize to produce $VCl^+$.

\begin{figure*}[htbp]
\vspace*{-0mm} \hspace{-2mm} {\hbox{\epsfxsize=140mm
\epsffile{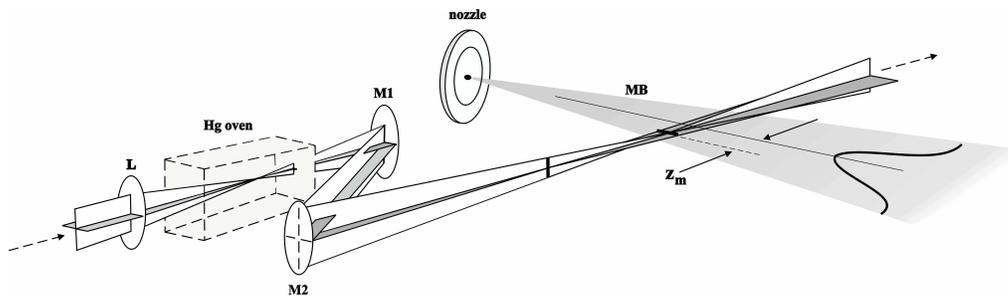}}} \vspace{0mm}
\caption{Schematic drawing of the apparatus. The 532 nm visible
laser is focused by a lens L ($f=76.2$ cm) into a mercury oven.
Mirrors M1 ($f=-5.1$ cm) and M2 ($f=7.6$ cm) are mounted inside the
$H_2$ phase tuning cell (not shown).  The two astigmatic foci are
separated by 4.6 mm.} \label{Apparatus}
\end{figure*}

    The experimental method is similar to that used previously \cite{apparatus paper}.
Key elements of the setup are depicted in Fig. \ref{Apparatus}.  A
pulsed nozzle beam of neat $VCl$ intersects the laser beams, which
are focused by a pair of mirrors, M1 and M2.  The molecular beam has
a Gaussian profile with a FWHM of 397 $\mu$m. The off-axis
configuration of the mirrors produces two astigmatic, elliptical
foci, one perpendicular to the plane defined by the laser and
molecular beam and the other in the plane. All the data reported
here used the in-plane (horizontal) focus. Mirror M2 is mounted on a
motorized stage, allowing the focal point to be translated across
the molecular beam with sub-micron resolution. The frequency of the
second harmonic of a Nd:YAG laser (532 nm) is tripled in a mercury
oven to produce 177 nm radiation, and the relative phase of the
fields ($\phi_3-3\phi_1$) is varied by passing the beams through a
chamber filled with hydrogen gas (not shown). The reaction products
are detected with a time-of-flight mass spectrometer. Additional
details will be provided in a future publication.

\begin{figure}[htbp]
\begin{center}
\vspace*{-5mm} \hspace{-10mm} {\hbox{\epsfxsize=70mm
\epsffile{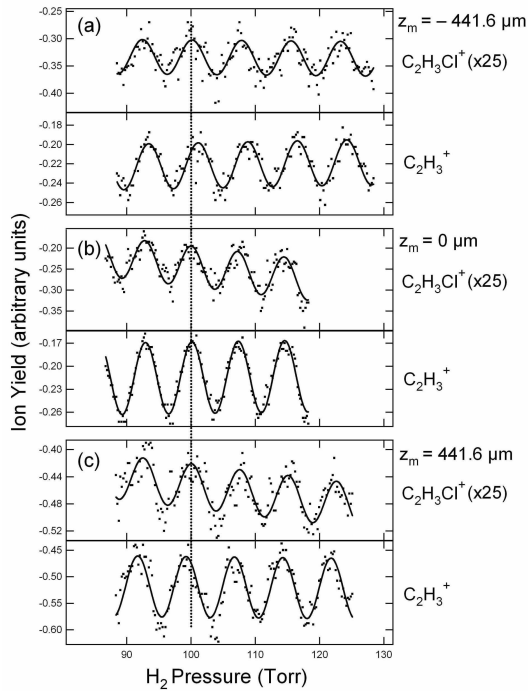}}} \vspace*{-5mm}
\end{center}
\caption{Modulation data for the parent ion and the $C_2H_3$
fragment, measured at (a) $z_m=-441.6 \mu m$, (b) $z_m=0$, and (c)
$z_m=+441.6 \mu m$. The pairs of modulation curves are shifted
horizontally so that the parent ion signals are all in phase.   The
solid curves are least squares sinusoidal fits to the data, and the
vertical dotted line is drawn to guide the eye.} \label{Modulation
data}
\end{figure}

\begin{figure}[htbp]
\begin{center}
\includegraphics[width=8cm]{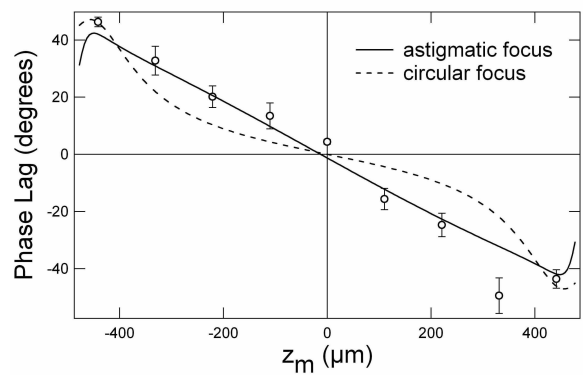}
\end{center}
\caption{Phase lag as a function of the distance of the molecular
beam axis from the focal line of the laser. A positive phase lag
corresponds to the parent ion signal leading the fragment. Error
bars for the data points at $z_m=0$ and $\pm441.6\mu m$ are the
standard deviations of repeated measurements, whereas the error bars
for the other points are derived from the least squares fits to a
single pair of modulation curves.  The dashed line is the analytical
result (Eq. (\ref{delta})) for a circular Gaussian focus and a
rectangular molecular beam profile. The solid curve is a numerical
calculation of the spatial phase, taking into account the astigmatic
focus of the laser beam and the Gaussian profile of the molecular
beam.} \label{Phase lags}
\end{figure}

Repeated scans of the molecular beam profile were recorded to
determine its peak location, which defines the origin of the z-axis.
Next, the ion yield vs. $H_2$ pressure (referred to as the
"modulation curve") was measured for $VCl^+$ and $C_2H_3^+$ for nine
positions of the laser focus, with the axis of the molecular beam
located at a distance $z_m$ from the focal line. Representative
modulation curves are shown in Fig. \ref{Modulation data}, recorded
on the axis of the molecular beam (panel (b)) and at the two extreme
positions (panels (a) and (c)). Repeated (3 to 6) measurements of
these curves obtained on different days yielded average phase lags
of $46.3\pm1.7^0$ at $z_m=-441.6\mu m$, $4.4\pm0.8^0$ at $z_m=0$,
and $-43.6\pm3.3^0$ at $z_m=+441.6\mu m$, where the uncertainty is a
single standard deviation for all the measurements at each point.
The least squares uncertainty of the fitted value of the phase lag
for a single scan is typically twice the standard deviation of the
mean for multiple  measurements at the same point. The phase lags at
all nine axial positions are plotted in Fig. \ref {Phase lags}.

A qualitative explanation of the spatial dependence of the phase lag
is as follows.  Although the Gouy phase shifts for the two channels
are identical at every point in space, the spatial distributions of
the product concentrations differ because of their different
intensity dependences.  A spatial average of the interference term
over the entire irradiated volume  yields  a net phase lag between
the products. A quantitative value of the spatial phase may be
obtained by averaging the interference term over the axial and
radial coordinates of the laser beam.  We assume for the moment a
circular Gaussian electric field,
\begin{eqnarray}
&&E(r,z)=E_0\frac{w_0}{w(z)} \\
&\times & \exp\left\{-i(\phi+kz-\eta(z))
-r^2\biggl[\frac{1}{w(z)^2}+\frac{ik}{2z\zeta(z)}\biggr]\right\}\, ,
\nonumber \label{Field}
\end{eqnarray}
where $E_0$ is the amplitude of the field, $w_0$ is the radius of
the focal spot, $w(z)=w_0\zeta(z)$ is the radius of the field at
axial distance $z$ from the focus, and $\zeta(z)=1+z^2/z_R^2$
describes the divergence of the beam \cite{Yariv}. We further assume
that $m$ visible and $n$ UV photons are absorbed in the control step
to produce the neutral fragments, and that $l$ additional visible
photons are absorbed to produce the parent ion.  We also assume that
the molecular beam has a rectangular profile of width $2d$. The
spatial average of the transition probability is then obtained by
inserting Eq. (4) into the off-diagonal (Eq. (\ref{Pmn})) and
diagonal matrix elements. Integrating first over $r$ and then over
$z$ yields the result
\begin{eqnarray}
&&\langle{P^S}\rangle\propto I_{l+n-1} + I_{l+m-1} \\
& +&
(2I_{l+(m+n)/2}-I_{l+(m+n)/2-1})\cos\bar{\phi}+2J_{l+(m+n)/2}\sin
\bar{\phi}, \label{crossterm} \nonumber \label{avarage}
\end{eqnarray}
where $\bar{\phi}=m\phi_n-n\phi_m+\delta_{mn}^S$, and the definite
integrals
\begin{equation}
I_n=\int_{-d-z_m}^{d-z_m}\frac{dz}{\zeta(z)^n},\hspace{2mm}
J_n=\int_{-d-z_m}^{d-z_m}\frac{z/z_R}{\zeta(z)^n}dz
\label{Integrals}
\end{equation}
have simple algebraic forms. Writing the cross term in the form
$2\langle P_{mn}^S\rangle \cos(\bar{\phi}+\phi_{sp})$, we obtain for
the spatial phase
\begin{equation}
\tan\phi_{sp}=-\frac{2J_{l+(m+n)/2}}{2I_{l+(m+n)/2}-I_{l+(m+n)/2-1}}.
\label{delta}
\end{equation}
The analytical value of $\phi_{sp}$  is given by the dashed curve in
Fig. \ref{Phase lags} for $d=400 \mu m$ and $z_R=64 \mu m$.  Even
better agreement with the data is obtained by taking into account
the astigmatism of the laser beam and the Gaussian profile of the
molecular beam.  A numerical evaluation of $\phi_{sp}$ is given by
the solid curve in Fig. \ref{Phase lags}, with no adjustable
parameters.  The small negative phase lag calculated for $z_m=0$ is
a residual effect of the second laser focus.  Details of the
calculation will be presented in a future publication.

The data in Fig. \ref{Modulation data} and the quantitative
agreement between experiment and theory in Fig. \ref{Phase lags}
provide a number of valuable insights.  First, our data demonstrate
that coherent phase modulation of large molecules is robust. Bersohn
$et$ $al.$ \cite{Bersohn} previously demonstrated control of
bound-to-bound transitions in polyatomic molecules, but here, and
also in ref. \cite{Tachiya}, it is shown that phase control of
reactive transitions in molecules having large densities of states
is achievable with modulation depths as large as $42\%$. Second, the
excellent agreement between experiment and theory is obtained only
for $m=3$, $n=1$, and $l=2$. This finding provides direct evidence
that the $C_2H_3^+$ signal is produced by photodissociation of the
neutral molecule at the three-photon level and not by fragmentation
of the parent ion.  In other words, we are controlling the branching
between ladder climbing and ladder switching. (If $C_2H_3^+$ was
produced by fragmentation of the parent ion, the Gouy phase lag
would vanish.) Of course, the branching ratio could also be
controlled by varying the total energy of the of the laser pulse,
but here we have shown that for a fixed set of laser conditions it
is possible to control the branching ratio $coherently$ by varying
only the relative phase of the two laser beams. Third, the positive
experimental phase lag at $z_m=0$ differs significantly (at the
$99.9\%$ level) from the theoretical value of $-1.3^0$.  We believe
that this phase lag is due to a molecular phase in one or both of
channels at the $3\omega_1$ level. The small value of the phase lag
is comparable in size to that found by Tachiya $et$ $al.$ for a
diffrerent molecule \cite{Tachiya}, but it is unclear at this point
whether larger channel phases might exist closer to the center of
the $\pi,\pi^*$ transition.

   It should be noted that the ionic wave function (at the
$5\omega_1$ level) does not contribute to the observed phase lag
because the two visible photons that connect the intermediate
$\pi,\pi^*$ state to the ionic state are present in the two
interfering paths ($\omega_3+2\omega_1$ and $5\omega_1$), so that
any molecular phase that is picked up in one path is exactly
cancelled by one in the other path.

In conclusion, we have shown that the Gouy phase of a focused laser
beam may be used to control the branching ratio of a photo-induced
reaction.  This phase, which was not included in previous
formulations of coherent phase control, adds linearly to the
refractive and molecular phases in the interference term. A
necessary and sufficient condition for the Gouy phase to serve as a
control parameter is that the product yields have different
intensity dependences.

We wish to thank the National Science Foundation for its generous
support under grant nos. PHY-0200812 and CHE-0120997.  Support by
the National Science Foundation of China under grant no. 10404008 is
acknowledged by ZH.

\bibliographystyle{phaip}

\end{document}